\documentclass[12pt,a4paper]{article}
\usepackage{amssymb,amsmath,amsthm,latexsym}
\usepackage{mathrsfs}
\usepackage[english]{babel}
\usepackage[latin1]{inputenc}
\usepackage{amsfonts}
\usepackage{graphics}
\usepackage{xcolor}

\def\x{\mathcal{X}}
\def\X{\mathcal{X}}

\def\fq{\mathbb{F}_q}
\def\bx{\boldsymbol{X}}
\def \C {\mathcal{C}}
\def\bd{\boldsymbol{\delta}}

\def\balpha{\boldsymbol{\alpha}}

\newtheorem{theorem}{Theorem}[section]
\newtheorem{corollary}[theorem]{Corollary}

\newtheorem{definition}[theorem]{Definition}

\begin{document}

\begin{center}
{\Large\textbf{A family of codes with variable locality and 
availability}}
\end{center}
\vspace{3ex}

\begin{center}

\textsc{C\'{\i}cero Carvalho and Victor G.L. Neumann}
\newline
\small{Faculdade de Matem\'{a}tica \\ Universidade Federal de 
Uberl\^{a}ndia
}
\end{center}

\vspace{8ex}

\vspace{4ex}
\begin{small}
\begin{center}
\textbf{Abstract}
\end{center}
In this work we present a class of locally recoverable codes, i.e.\ codes where 
an erasure at a position $P$ of a codeword may be recovered from the knowledge 
of the entries in the positions of a recovery set $R_P$. The codes in the class 
that we define have availability, meaning that for each position $P$ there are 
several distinct recovery sets. Also, the entry at position $P$ may be 
recovered 
even in the presence of erasures in some of the positions of the recovery sets, 
and the number of supported erasures may vary among the various recovery sets.
\end{small}

\section{Introduction}

Motivated by applications in large scale distributed storage systems, Gopalan 
et.\ al.\ 
(\cite{gopalan}) introduced the concept of locally recoverable codes, in order 
to recover an erasure in a codeword (due to a node failure in such a storage 
system, for example). The idea is to attach to each position $j$ in a codeword 
a recovering set $R_j$ of at most $r \geq 1$ other positions such that  the 
entry at 
position $j$ may be recovered from the entries at the positions in the set 
$R_j$. Such a code is said to have {\em locality $r$}. In \cite{prakash} 
Prakash et.\ al.\  modified this idea, to take into account 
possible erasures in the recovering set. The definition is as follows.

\begin{definition}  \label{1st-def}
Let $r$ and $\delta$ be positive integers, with $\delta \geq 2$. A linear code 
$\C$  of length $m$  is said to have locality $(r, \delta)$ if for every 
position $j \in \{1, \ldots, m\}$ there exists a subset $S_j \subset \{1, 
\ldots, m\}$ such that: \\
a) $j \in S_j$ and $| S_j | \leq r + \delta - 1$; \\
b) the minimum distance of the code $\C |_{S_j}$, obtained from $\C$ by 
considering only the entries in the positions of $S_j$, is at least $\delta$.   
\end{definition}

From the definition we get that two distinct codewords of $\C |_{S_j}$ can 
coincide in at most $r-1$ positions (otherwise there would exist a nonzero 
codeword in  $\C |_{S_j}$ with weight less than $\delta$), so we may conclude 
that the entries in any set of $r$ positions in $S_j$ determine the entries in 
the remaining $\delta - 1$ positions. The idea is that an erasure in the 
position $j$ in a codeword of $\C$ may be recovered from the entries in the 
positions of $S_j$, 
even in the presence 
of $\delta - 2$ erasures in positions of $S_j \setminus \{ j \}$.

More recently Rawat et.\ al.\ extended the original idea of locally recoverable 
codes in another direction. In \cite{rawat} they introduced the concept of 
locally recoverable codes {\em with availability $n$}, which are codes where 
to each position $j$ in a codeword are associated 
$n$ pairwise disjoint recovering sets $R_{1,j}, \ldots, R_{n,j}$, each with 
at most $r \geq 1$ other positions,  such that  the 
entry at 
position $j$ may be recovered from the entries at the positions in any set 
$R_{i, j}$, with $i \in \{1, \ldots, n\}$.

In this paper we present a construction which combines both features described 
above. In Section 1 we show how to produce codes with availability 
$n$, with the difference that, for certain positive integers $\delta_1, \ldots, 
\delta_n$ and $r_1, \ldots, r_n$ the set $R_{i, j}$ contains position $j$, has 
cardinality equal to $r_i + \delta_i - 1$, and the entries any set of $r_i$ 
positions in $R_{i, j}$ determine the entries in the remaining $\delta_i - 1$
positions of $R_{i, j}$, for $i = 1, \ldots, n$ (see Theorem \ref{main}).
Since the integers $r_1, \ldots, r_n$ need not to coincide, we say that these 
codes have {\em variable locality}.
We also determine the dimension and the minimum distance of such codes. In the 
last section we present some numerical results on this class of codes. 

\section{Main results}

Let $\fq$ be a finite field with $q$ elements. 
For $i  = 1, \ldots , n$ let $A_i \subset \fq$, with $| A_i| =: d_i$ and we 
will assume that $2 \leq d_1  \leq \cdots \leq d_n$.
Let $\x := A_1 \times \cdots \times A_n$.

For $i  = 1, \ldots , n$ let $f_i = \prod_{a_i \in A_i} (X_i - a_i) \in 
\fq[X_1, \ldots, X_n] =: \fq[\bx]$, it is well known (see e.g. \cite[Lemma 
2.3]{lrv}) that the ideal consisting of all polynomials in $\fq[\bx]$ which 
vanish on all points of $\x$ is  
$I_{\x} := (f_1, \ldots , f_n)$.

Let $m := d_1 . \cdots . d_n$ and write $\x = \{P_1, \ldots, P_m\}$.  Let  
\[
\varphi: \fq[\bx]/I_\x \rightarrow \fq^m
\]
 be the evaluation morphism defined 
by $\varphi(f  + I_\x) = ( f(P_1), \ldots, f(P_m))$. Clearly $\varphi$ is 
injective, and it is also surjective because for each $P_i \in \x$ one may find 
$g_i \in \fq[\bx]$ such that $g_i(P_i) = 1$ and $g_i(P_j) = 0$ for $j \in \{1, 
\ldots, n\} \setminus \{i\}$.  

\noindent
For all $i \in \{1, \ldots , n\}$ let $\delta_i \geq 2$ be an integer such that 
$d_i - \delta_i + 1 =: r_i \geq 1$. Let $d$ be a positive integer and let $L_d 
\subset \fq[\bx]/I_\x $ be the set 
\[
L_d = \{ f + I_\x \mid \deg(f) \leq d \textrm{ and } \deg_{X_i}(f) < r_i 
\textrm{ 
for all } i = 1, \ldots, n\} \cup \{ 0 + I_\x \}.
\]
Observe 
that $L_d$ is an $\fq$-vector space so the image $\varphi(L_d)$ is a linear 
code of 
length $m$ (see \cite{fitzgerald-lax} for results on the decoding of such 
codes). We write $\bd := 
\{\delta_1, \ldots, \delta_n\}$ and define 
\[
\C(\x, d, \bd) := \varphi(L_d).
\]

\begin{theorem} Let $j \in \{1, \ldots, m\}$ and let $\boldsymbol{c} \in \C(\x, 
d, 
\bd)$. For every $i 
\in \{1, \ldots, n\}$ there is a set $R(i,j)$ of positions in the codeword 
$\boldsymbol{c}$ (or, 
equivalently, a set of points of $\x$), such that: \\
a) $j \in R(i,j)$; \\
b) $|R(i,j)| = d_i$; \\
c) the entry at the $j$-th position of $\boldsymbol{c}$ may be recovered from 
the knowledge of the entries at any set of $r_i$ positions in $R(i,j) \setminus 
\{j\}$;\\
d) $R(i_1,j) \cap R(i_2,j) = \{ j \}$ for any distinct $i_1, i_2 \in \{1, 
\ldots, n\}$.
\end{theorem}
\begin{proof}
Let $f$ be a polynomial of degree at most $d$,  with $\deg_{X_i}(f) < 
r_i $ for all $i = 1, \ldots, n$, and let $P \in \x$ be the point such that the 
$j$-th entry of $\boldsymbol{c}$ is $f(P)$. Let $i \in \{1, 
\ldots, n\}$, write $P = 
(\alpha_1, \ldots, \alpha_n)$ and let 
\[
R(i,j):= \{ (\alpha_1, \ldots, \alpha_{i - 1}, \beta, \alpha_{i + 1}, \ldots,   
\alpha_n) \in \x \mid \beta \in A_i\}.
\]
Clearly $|R(i,j)| = d_i$, $P \in R(i,j)$ and $R(i_1,j) \cap R(i_2,j) = \{ P \}$ 
for any distinct $i_1, i_2 \in \{1, 
\ldots, n\}$.  Write $f = \sum_{t = 0}^{r_i - 
1} 
g_t X_i^t$, where we have $g_t \in \fq[X_1, \ldots, X_{i - 1}, X_{i + 1}, 
\ldots, X_n]$ 
for all $t = 0, \ldots, r_i - 1$. 
Observe that, given $t \in \{0, \ldots, r - 1\}$ the value of $g_t$ at any 
point of $R(i,j)$ is constant, say $e_t$. 
Assume that there are $r_i$  distinct points in $R(i,j) \setminus\{ P \}$, say 
$Q_s = (\alpha_1, 
\ldots, \alpha_{i - 1}, \beta_s, \alpha_{i + 1}, \ldots,   
\alpha_n)$, for which we know the 
value $c_s := f(Q_s)$, where $s = 1, \ldots, r_i$. From this set of equalities 
we may determine the values of $e_0, \ldots, e_{r_i - 1}$ by solving the system
\[
\left(
\begin{matrix}
1 & \beta_1 & \beta_1^2 & \cdots & \beta_1^{r-1} \\
1 & \beta_2 & \beta_2^2 & \cdots & \beta_2^{r-1} \\
1 & \beta_3 & \beta_3^2 & \cdots & \beta_3^{r-1} \\
1 & \colon & \colon & \ddots & \colon \\
1 & \beta_r & \beta_r^2 & \cdots & \beta_r^{r-1} 
\end{matrix}
\right)
\left(
\begin{matrix}
e_0 \\
e_1 \\
e_2\\
\colon \\
e_{r-1}
\end{matrix}
\right)=
\left(
\begin{matrix}
c_1 \\
c_2 \\
c_3 \\
\colon \\
c_{r}
\end{matrix}
\right)
\]
which has a unique solution because $\beta_u \neq \beta_v$ for distinct $u$ and 
$v$ in $\{1, \ldots, r\}$.

If $Q  = (\alpha_1, \ldots, \alpha_{i - 1}, \beta, \alpha_{i + 1}, \ldots,   
\alpha_n) \in R(i,j)$ then $f(Q) =  \sum_{t = 0}^{r_i - 1}e_t \beta^t$ so 
with the knowledge of $e_0, \ldots, e_{r_i - 1}$ we determine $f(Q)$ for 
any $Q \in R(i,j)$.
\end{proof}

\noindent
Now we want to determine the dimension of $\C(\x, d, \bd)$.
From the definition of $\C(\x, d, \bd)$ we get that $d \leq \sum_{i = 1}^n (r_i 
- 1)$.

\begin{theorem} \label{main} The dimension of $\C(\x, d, \bd)$ for $0 \leq d 
\leq \sum_{i = 
1}^n (r_i - 1)$ is given by 
\begin{equation*}
\begin{split}
   \dim(\C(\x, d, \bd)) =  & \binom{n + d }{d}  -   \sum_{i = 1}^n \binom{n + d 
  - r_i}{d - 
  r_i} + \cdots +   \\ & (-1)^j \sum_{1 \leq i_1 < \cdots < i_j \leq n} 
  \binom{n + d - r_{i_1} - \cdots -  r_{i_j}}{d - r_{i_1} - \cdots -   r_{i_j}} 
  + \cdots +  \\ & (-1)^n \binom{n + d - r_{1} - \cdots -   r_{n}}{d - r_{1} - 
  \cdots -   r_{n}} 
\end{split} 
\end{equation*}
where we set $\binom{a}{b} = 0$ if $b < 0$.
\end{theorem}
\begin{proof}
It is known that the basis $\{f_1, \ldots, f_n\}$ for $I_{\x}$ is a Gr\"obner 
basis (with 
respect to any monomial order) so any set of classes of monomials of 
the form 
$X_1^{m_1} . \cdots . X_n^{m_n} + I_{\x}$, with $0 \leq m_i \leq d_i - 1$ for 
all $i = 
1, \ldots, n$, is linearly independent over $\fq$. In particular the set 
\begin{equation} \label{B}
B := 
\{ 
X_1^{m_1} . \cdots . X_n^{m_n} + I_{\x} \mid  0 \leq m_i \leq r_i - 1, \, i = 
1, \ldots, n, \sum_{j = 1}^n m_j \leq d\},
\end{equation}
whose image under $\varphi$ generates $\C(\x, d, \bd)$, is 
linearly independent. From the fact that $\varphi$ is injective, we get that 
$\dim \C(\x, d, \bd) = | B |$. 

To count the number of elements in $B$ we observe that for each monomial 
$X_1^{m_1} . \cdots . X_n^{m_n} \in B$ there corresponds a unique monomial in 
$\fq[X_0, X_1, \ldots, X_n]$  of 
degree d, namely  $X_0^{d - \sum_j m_j} .X_1^{m_1} . \cdots . X_n^{m_n}$. Thus 
the number $| B |$ is precisely the coefficient of $t^d$ in the series  
\[
 H(t) := (1 + t + t^2 + \cdots ) . (1 +  t + \cdots + t^{r_1 - 1}) . \cdots . 
 (1 +  t + \cdots + t^{r_n - 1}).
\]
We may think of $H(t)$ as a real function of one variable $t$ defined in a 
suitable neighborhood of $0$, say $| t | < 1$, so that $1 + t + t^2 + \cdots = 
1/(1 - t)$ and
$$ H(t) = \frac{1}{1 - t} .  \frac{1 - t^{r_1}}{1 - t} . \cdots . \frac{1 - 
t^{r_n}}{1 - t}.
$$
Thus we get $H(t) = ( 1/(1 - t)^{n + 1} )  \prod_{i = 1}^n (1 - t^{r_i})$, 
and from 
$1/(1 - t)^{n + 1} = \sum_{j = 0}^\infty \binom{n + j}{j} t^j$ we get
\begin{equation*}
\begin{split}
H(t) =  (\sum_{j = 0}^\infty \binom{n + j}{j} t^j)&\left(1 - \sum_{i = 1}^n 
t^{r_i}  
+ \sum_{1 \leq i_1 < i_2 \leq n} t^{r_{i_1} + r_{i_2}} + \cdots + \right. \\ &
\left. (-1)^j \sum_{1 \leq i_1 < \cdots < i_j \leq n} t^{r_{i_1} +  \cdots + 
r_{i_j}}
+ \cdots + 
(-1)^n t^{r_{i_1} +  \cdots + r_{i_n}}\right).
\end{split}
\end{equation*}
The expression for $\dim \C(\x, d, \bd)$ in the statement of the theorem is the 
coefficient of $t^d$ in $H(t)$, obtained from the above product.
\end{proof}

\noindent
We recall some notation from Algebraic Geometry and some facts on varieties 
defined over finite fields and from Gr\"obner basis theory.
Given an ideal $I \subset \fq[\bx]$ we denote by $V(I) \subset \fq^n$ the set 
of $n$-tuples which vanish on all polynomials of $I$. From now on fix a 
monomial order in the set of monomials of $\fq[\bx]$. We denote by $\Delta(I)$ 
the set, sometimes called the {\em footprint} of $I$,  of monomials in 
$\fq[\bx]$ which do not appear as leading monomials 
of polynomials in $I$. Given a set $M_1, \ldots, M_s$ of monomials 
in $\fq[\bx]$ we denote by $\Delta(M_1, \ldots, M_s)$ the set of monomials 
which are not multiple of any $M_i$, for $i = 1, \ldots, s$. In what follows we 
will need the following result.

\begin{theorem}\label{variedade-pegada}\cite[Lemma 6.50 and Thm. 8.32]{becker} 
Let $I \subset \fq[\bx]$ be an ideal 
such that $\Delta(I)$ is a 
finite set and  let $L$ be an 
algebraically closed extension of $\fq$. Then $V_L(I) := \{(a_1, \ldots, a_n) 
\in 
L^n \; | \; f(a_1, \ldots, a_n)= 0 \textrm{ for all } f \in I\}$ is a finite 
set and $| V_L(I))| \leq | \Delta(I)|$. Moreover, if  $I$ is a radical ideal 
then 
$| V_L(I) | = | \Delta(I) |$.
\end{theorem} 

\noindent
Now we determine the minimum distance of $\C(\x, d, \bd)$.

\begin{theorem} Let $m(s_1, \ldots , s_n) :=  \prod_{i = 1}^n (d_i - s_i)$
where $0 \leq s_i \leq r_i - 1$ is  an integer for all 
$i = 1,\ldots, n$, and let $d$ be an integer such that $0 \leq d \leq 
\displaystyle \sum_{i = 1}^n (r_i - 1)$.  Then the minimum distance of $\C(\x, 
d, \bd)$ is equal to $\mu$ where
\[
\mu = \min \{ m(s_1, \ldots, s_n) \, | \, s_1 + \cdots + s_n \leq d \}.
\]
\end{theorem}
\begin{proof} Let $f$ be a polynomial such that 
$\deg(f) \leq d$ and $\deg_{X_i}(f) \leq r_i - 1 = d_i - \delta_i $ for all $i 
= 1, 
\ldots, n$ and let $J_f := (f, f_1, \ldots, f_n)$. Then the number of zeros in 
the 
codeword $\varphi(f + I_\x)$ is equal to $| V(J_f) |$ so that the weight of 
this 
codeword is $\omega(\varphi(f + I)) = \prod_{i = 1}^n d_i - | V(J_f) |$.
If $L$ is an algebraically closed extension of $\fq$ then $V(J_f) \subset  
V_L(J_f)$. On the other hand 
\[
\Delta(J_f) \subset \Delta(I_\X) \subset 
\{ 
X_1^{u_1} . \cdots . X_n^{u_n} \mid  0 \leq u_i \leq d_i - 1, \, i = 
1, \ldots, n \}.
\]  
Thus we can apply  
Theorem \ref{variedade-pegada} and we  get that  $| V(J_f) | \leq | \Delta(J_f) 
|$. 
Let 
$M := 
X_1^{m_1} . \cdots . X_n^{m_n}$
be the leading monomial of $f$. From the definition of footprint and  
properties of monomial orders we get that 
$\Delta(J_f) 
\subset \Delta(M, X_1^{d_1}, \ldots, X_n^{d_n})$
so that $| \Delta(J_f) | 
\leq   
\prod_{i = 1}^n d_i - \prod_{i = 1}^n (d_i - m_i)$. Thus 
\[
\omega(\varphi(f + 
I_\x))\geq 
\prod_{i = 1}^n (d_i - m_i) \geq \mu.
\]
Let $0 \leq t_i \leq r_i - 1$, for $i= 1, \ldots, n$ be such that $\sum_{i = 
1}^nt_i \leq d$ and
$\prod_{i = 1}^n (d_i - t_i) = \mu$, and write 
$A_i := \{ a_{i 1}, \ldots, a_{i 
d_i} \}$ for $i = 1, \ldots, n$. Let 
\[
g(X_1, \ldots, X_n) =  \prod_{i = 
1}^n \prod_{j = 1}^{t_i} (X_i - a_{i j}), 
\]
 then $\deg_{X_i}(g) = t_i \leq r_i - 1$ for all $i = 1, 
 \ldots, n$, $\deg(g)  \leq d$,  
and clearly $g$ has exactly $\mu$ non-zeros in $\x$.
\end{proof}

We present a formula for the minimum distance in a special case.

\begin{theorem}
Assume that $2 \leq d_1 \leq \cdots \leq d_n$ and that $\delta_1 \leq 
\cdots \leq \delta_n$. Then the minimum distance of $\C(\x, d, \bd)$ for 
$0 \leq d \leq 
\sum_{i = 1}^n (r_i - 1)$ is given by 
\[
\delta_1. \cdots .\delta_k (d_{k + 1} - \ell) \prod_{i = k + 2}^n d_i,
\] 
where $k$ and $\ell$ are uniquely determined  by writing $d = \sum_{i = 1}^k 
(d_i - \delta_i) + 
\ell$ with $0 \leq \ell < d_{k + 1} - \delta_{k+1}$. 
\end{theorem}
\begin{proof}
From the above theorem it suffices to prove that 
the minimum value of  $m(s_1, \ldots , s_n) =  \prod_{i = 1}^n 
(d_i - s_i)$ where $0 \leq s_i \leq d_i - \delta_i$ for all $i = 1, \ldots, n$ 
and $\sum_{i = 1}^n s_i \leq d$ 
is attained at the $n$-tuple $(d_1 - \delta_1, \ldots,  d_k - \delta_k, \ell, 
0, \ldots, 0)$. Clearly the minimum of $m(s_1, \ldots , s_n)$ is attained when 
$\sum_{i = 1}^n s_i = d$, so we assume that. 

Suppose that there exists an index, say $i_1$, which is the least such that 
$i_1 \leq k$ and $s_{i_1} < d_{i_1} - \delta_{i_1}$. If 
there exists $i_2 \in \{1, \ldots , n\}$ such that $i_1 < i_2$, $s_{i_2} > 0$  
and $s_{i_1} + s_{i_2} \leq d_{i_1} - \delta_{i_1}$, then denoting by $(s'_1, 
\ldots , s'_n)$  the 
$n$-tuple obtained from $(s_1, \ldots , s_n)$ by replacing $s_{i_1}$ by 
$s_{i_1} + s_{i_2}$ and 
$s_{i_2}$ by 0, we get that 
\[
m(s_1, \ldots , s_n) - m(s'_1, 
\ldots , s'_n) = (s_{i_1} s_{i_2} + (d_{i_2} - d_{i_1})s_{i_2})  
\prod_{\stackrel{i = 1}{i\neq i_1, \, i_2}}^n (d_i - s_i) \geq 0
\] 	
so that $m(s'_1, \ldots , s'_n) \leq m(s_1, \ldots , s_n)$, and note that 
the strict inequality holds if $s_{i_1} > 
0$ or $d_{i_2} > d_{i_1}$. In the case there exists $i_2 \in \{1, \ldots , n\}$ 
such that $i_1 < i_2$, 
$s_{i_2} > 0$  and $s_{i_1} + s_{i_2} > d_{i_1} - \delta_{i_1}$, then denoting 
by $(s''_1, \ldots , s''_n)$   the 
$n$-tuple obtained from $(s_1, \ldots , s_n)$ by replacing $s_{i_1}$ by 
$d_{i_1} - \delta_{i_1}$ and 
$s_{i_2}$ by $s_{i_2} - (d_{i_1} - \delta_{i_1} - s_{i_1})$ we get that 
\[
m(s_1, \ldots , s_n) - m(s''_1, \ldots , s''_n) = (d_{i_1} - \delta_{i_1} - 
s_{i_1})(d_{i_2} - \delta_{i_1} - s_{i_2}) 
\prod_{\stackrel{i = 1}{i\neq i_1, \, i_2}}^n (d_i - s_i) .
\]
From the hypothesis we get that $d_{i_2} - \delta_{i_2} \leq  d_{i_2} - 
\delta_{i_1}$ so $d_{i_2} - \delta_{i_1} - s_{i_2} \geq 0$ and a fortiori
$m(s_1, \ldots , s_n) \geq m(s''_1, \ldots , s''_n)$.
This proves that, starting with an $n$-tuple $\boldsymbol{s}$ we can find an 
$n$-tuple 
$\boldsymbol{s'}$ such that $m(\boldsymbol{s'}) \leq m(\boldsymbol{s})$ and the 
first $k$ entries of $\boldsymbol{s'}$ are 
equal to $d_1 - \delta_1, \ldots, d_k - \delta_k$, in that order. By a similar 
reasoning we prove 
that if $\ell > 0$ then the minimum is attained when the $(k+1)$-th entry of 
$\boldsymbol{s'}$ is equal to 
$\ell$ (and 
the other entries are necessarily equal to zero).

\end{proof}

\section{MDS codes and numerical examples}

In this session we examine more closely the restriction of the code 
$\C(\x, d, \bd)$  to sets of the form $R(i,j)$ which appeared in Theorem 
\ref{main}. 
In that Theorem we started by fixing an arbitrary position $j$ in the codeword 
to show that $\C(\x, d, \bd)$ is a locally recoverable code with availability 
$n$. It is implicit in the proof that the restriction of  
$\C(\x, d, \bd)$ to the positions in $R(i,j)$ is a code with locality $(r_i, 
\delta_i)$, for all $i = 1, \ldots, n$. To study this restriction the position 
$j$ is not so important anymore, so we modify the notation for the set $R(i,j)$.

Let $i \in \{1, \ldots, n\}$ and let $\balpha \in A_1 \times \cdots \times 
\widehat{A_i} \times \cdots \times A_n$, where $\widehat{A_i}$ means that the 
set $A_i$ is not a factor of the cartesian product. Let 
\[
S(\balpha, i) := \{ (\alpha_1, \ldots, \alpha_{i - 1}, \beta, \alpha_{i + 1}, 
\ldots,   
\alpha_n) \in \x \mid \beta \in A_i\}
\]
(so that $S(\balpha, i)$ coincide with $R(i,j)$ for certain values of $j$). We 
will denote by $\C(d, \balpha, i)$ the restriction of $\C(\x, d, \bd)$ to the 
positions corresponding to all points in $S(\balpha, i)$, so 
$\C(d, \balpha, i)$ is a code of length $d_i = |S(\balpha, i)| = | A_i |$.

\begin{theorem} \label{data-c-alpha}
Let $i \in \{1, \ldots, n\}$, let  $I_i = ( \prod_{\beta \in A_i} (X_i - \beta) 
) 
\subset \mathbb{F}_q[X_i]$, and let 
\[
\varphi_i : \mathbb{F}_q[X_i]/I_i \rightarrow 
\mathbb{F}_q^{d_i}
\]
be defined by $\varphi_i( f + I_i) = ( f(\beta_1), \ldots, f(\beta_{d_i}) )$.
Let 
\[ T_d =  \{ f + I_i \mid \deg(f) \leq \min\{d, r_i - 1\} \} \cup \{ 0 + I_\x 
\}.
\]
If $d < r_i - 1$ 
then the code $\C(d, \balpha, i)$ is equivalent to $\varphi_i(T_d)$, and 
if $d \geq r_i - 1$ then $\C(d, \balpha, i)$ is equivalent to $\varphi_i(T_{r_i 
-1})$. In particular, we have
\[
\dim(\C(d, \balpha, i)) = \left\{ \begin{array}{ll} d + 1 & \textrm{ if } d < 
r_i - 1 \\ r_i & \textrm{ if } d \geq r_i - 1 \end{array} \right.;
\]
and for the minimum distance $\delta(\C(d, \balpha, i))$ of $\C(d, \balpha, i)$ 
we have 
\[ \begin{array}{ll}
\delta(\C(d, \balpha, i)) =  d_i - d, & \textrm{ if } d < r_i - 1; \\
\delta(\C(d, \balpha, i)) =  d_i - r_i + 1 \; (= \delta_i), & \textrm{ if } d 
\geq 
r_i - 1.
\end{array}
\]
\end{theorem}
\begin{proof}
For all $j \in \{1, \ldots, d_i\}$ let $Q_j =  
(\alpha_1, \ldots, \alpha_{i - 1}, \beta_j, \alpha_{i + 1}, \ldots,   
\alpha_n)$, 
so that $S(\balpha, i)  = \{Q_1, \ldots, Q_{d_i}\}$.
The $d_i$-tuples which form 
restriction of $\C(\x, d, \bd)$ 
to the 
positions corresponding to the points in $S(\balpha, i)$ are, except possibly 
for the order of the coordinates, exactly the $d_i$-tuples in the 
image 
$\widetilde{\varphi_i}(L_d)$, where $\widetilde{\varphi_i}: \fq[\bx]/I_\x 
\rightarrow \fq^{d_i}$ is
the linear transformation defined by $\widetilde{\varphi_i}( f + I_\x) = ( 
f(Q_1), \ldots, f(Q_{d_i}) )$.  
Set $e_i = \min\{d, r_i - 1\}$,  we know that 
$\{1 + I_i, X_i + I_i, \ldots, X_i^{e_i} + I_i\}$ is a basis 
for $T_d$, and from \eqref{B} we get that
$\{1 + I_\X, X_i + I_\X, \ldots, X_i^{e_i} + I_\X\} \subset B$
is also linearly independent, so   
$\varphi_i(T_d) \subset \widetilde{\varphi_i}(L_d)$. On the other hand, given 
$f + I_\X \in L_d$ clearly we have that $f(\alpha_1, \ldots, \alpha_{i - 1}, 
X_i, \alpha_{i+1}, \ldots, \alpha_n) + I_i \in T_d$ and
\[
\widetilde{\varphi_i}( f + I_\x) = \varphi_i(f(\alpha_1, \ldots, \alpha_{i - 
1}, X_i, \alpha_{i+1}, \ldots, \alpha_n) + I_i)
\]
so that $\widetilde{\varphi_i}(L_d) \subset \varphi_i(T_d)$. This shows that 
$\C(d, \balpha, i)$ is equivalent to $\varphi_i(T_d)$, and we observe that if 
$d \geq r_i - 1$ then $T_d = T_{r_i - 1}$.

We know that $\{1 + I_i, X_i + I_i, \ldots, X_i^{d_i - 1} + I_i\}$ is a basis 
for 
$\mathbb{F}_q[X_i]/I_i$ as an $\mathbb{F}_q$-vector space and clearly 
$\varphi_i$ is an injective transformation, so $\varphi_i$ is an isomorphism. 
Thus $\dim(\C(d, \balpha, i)) = \dim(T_d)$, and if $d < r_i - 1$ then 
$\dim(T_d) = d + 1$, while if $d \geq r_i - 1$ then 
$\dim(T_d) = \dim(T_{r_i - 1}) = r_i$.

Observe that $e_i = \min\{d, r_i - 1\} < d_i$ and given $e_i$ distinct points 
in $S(\balpha, i)$ there exists $f + I_i \in T_d$ having exactly these points 
as zeros. Also, $e_i$ is the maximum number of points of $S(\balpha, i)$ which 
annihilate any nonzero class $f + I_i \in T_d$. This, and the fact that 
$T_d = T_{r_i - 1}$ prove the results on the minimum distance.
\end{proof}

The Singleton inequality states that for a code of length $m$, dimension $k$ 
and minimum distance $\delta$ one has $\delta + k \leq m + 1$. The codes for 
which equality holds are called {maximum distance separable}, or just MDS 
codes, and they have an optimal balance between the dimension and the minimum 
distance for the given length.

\begin{corollary}
For any $i \in \{1, \ldots, n\}$ the code $\C(d, \balpha, i)$ is an MDS code.
\end{corollary} 
\begin{proof}
This is a straight forward verification, using the data from the previous 
result.
\end{proof}

We finish this  section with numerical data obtained from the 
above results. In the tables, 
we use the following notation: $m=|\X|$ is the 
length of $\C(\x, d, \bd)$, $\kappa = \dim_{\mathbb{F}_q} \C(\x, d, \bd)$, 
and $v$ is the minimum distance of $\C(\x, d, \bd)$.


\begin{table}[htbp]
	\centering
	\caption{$\X := \mathbb{F}_{11} \times \mathbb{F}_{11} \times \mathbb{F}_{11}$}
	\begin{tabular}{c|cccccccccc}
		\multicolumn{11}{c}{$\bd :=  \{4, 5 , 6\}$} \\
		\hline
		$d$  & $4$ & $5$   &$8$ &$10$&$12$&$14$&$15$&$16$ & $17$ & $18$  \\
		\hline
		$m$  & $1331$& $1331$&$1331$&$1331$&$1331$&$1331$&$1331$&$1331$&$1331$&$1331$ \\
		\hline
		$\kappa$&$35$&$56$&$150$&$221$&$280$&$316$&$326$&$332$&$335$&$336$ \\
		\hline
		$v$    & $847$& $726$&$440$   & $352$  &$264$  & $200$&$180$&  $160$&$140$&$120$
	\end{tabular}
	\label{caseq749}
\end{table}

\begin{table}[htbp]
	\centering
	\caption{$\X :=   \mathbb{F}_{25} \times \mathbb{F}_{25}$}
	\begin{tabular}{c|cccccccccc}
		\multicolumn{11}{c}{$\bd :=  \{6,7\}$} \\
		\hline
		$d$  &  $5$   &$6$&$14$&$15$&$26$&$27$ & $28$ & $35$ & $36$  & $37$ \\
		\hline
		$m$  & $625$& $625$&$625$& $625$ &  $625$&$625$&$625$&$625$&$625$  &$625$ \\
		\hline
		$\kappa$&$21$&$28$& $120$&$136$&$314$&$325$& $335$ &$377$&$379$  & $380$ \\
		\hline
		$v$    & $500$&$475$&   $275$&$250$ &$108$&$102$&  $96$&$54$&$48$ & $42$
	\end{tabular}

	\label{caseq52525}
\end{table}


\end{document}